\documentclass[journal]{IEEEtran}
\usepackage[left=0.75in, right=0.75in, top=1in, bottom=0.9in]{geometry}
\makeatletter
\def\ps@headings{%
\def\@oddhead{\mbox{}\scriptsize\rightmark \hfil \thepage}%
\def\@evenhead{\scriptsize\thepage \hfil \leftmark\mbox{}}%
\def\@oddfoot{}%
\def\@evenfoot{}}
\makeatother 

\usepackage{amsmath,amssymb,mathrsfs,amsthm}
\usepackage{mathtools}
\usepackage{cite}

\usepackage{graphicx,stfloats,subfigure,multicol}%amsmath
\usepackage{epsfig,epsf,psfrag,amssymb,amsfonts,latexsym,graphicx,mathrsfs,subfigure}

\usepackage{comment}

\usepackage{booktabs}
\usepackage[table]{xcolor}
\usepackage{multirow}
\usepackage{tablefootnote}
\usepackage{bbm}
\usepackage{chngpage}
\usepackage{textcomp}
\usepackage{hyperref}
\usepackage{url}
\usepackage[hyphenbreaks]{breakurl}
\usepackage[normalem]{ulem}
\usepackage{algpseudocode}
\newsavebox{\ieeealgbox}

\usepackage{array}

\newtheorem{theorem}{Theorem}
\newtheorem{proposition}{Proposition}

\newtheorem{lemma}{Lemma}

 \def\old#1{}

\def\nn{\nonumber}
\def\beq{\begin{equation}}
\def\eeq{\end{equation}}
\def\bea{\begin{eqnarray}}
\def\eea{\end{eqnarray}}
\def\ba{\begin{array}}
\def\ea{\end{array}}

\def\bitem{\begin{itemize}}
\def\eitem{\end{itemize}}
\def\ben{\begin{enumerate}}
\def\een{\end{enumerate}}

\def\eg{{\it e.g., \/}}

\definecolor{bgrd}{rgb}{1,1,1}
\definecolor{gray}{rgb}{0.5,0.5,0.5}
\definecolor{dkr}{rgb}{0.7,0.1,0.2}
\definecolor{dkb}{rgb}{0.1,0.1,0.8}

\def\tcb{\textcolor{blue}}

\def\scalefig#1{\epsfxsize #1\textwidth}

\def\T{\mbox{\tiny T}}

\newcommand{\mbbE}{\mathbb{E}}

\newcommand{\Rmsc}{\mathscr{R}}

\newcommand{\Zmsc}{\mathscr{Z}}

\def\dbf{{\bf d}}

\def\Ec{{\cal E}}

\def\Lc{{\cal L}}

\def\Pc{{\cal P}}

\begin{document}

\title{On Net Energy Metering X: Optimal Prosumer Decisions, Social Welfare, and Cross-Subsidies}

\author{Ahmed S. Alahmed,~\IEEEmembership{Student~Member,~IEEE},
Lang~Tong,~\IEEEmembership{Fellow,~IEEE},
\thanks{\scriptsize  Ahmed S. Alahmed and
Lang Tong  ({\tt \{\tcb{ASA278,~LT35}\}\tcb{@cornell.edu}}) are  with the School of Electrical and Computer Engineering, Cornell University, USA.
This work was supported in part by the National Science Foundation under Grants 1932501 and 1816397.}}
\maketitle

{
\begin{abstract}
We introduce NEM X, an inclusive retail tariff model that captures features of existing net energy metering (NEM) policies. It is shown that the optimal prosumer decision has three modes: (a) the net-consuming mode where the prosumer consumes more than its behind-the-meter distributed energy resource (DER) production when the DER production is below a predetermined lower threshold, (b) the net-producing mode where the prosumer consumes less than its DER production when the DER production is above a predetermined upper threshold, and (c) the net-zero energy mode where the prosumer's consumption matches to its DER generation when its DER production is between the lower and upper thresholds.  Both thresholds are obtained in closed-form. Next, we analyze the regulator's rate-setting process that determines NEM X parameters such as retail/sell rates, fixed charges, and price differentials in time-of-use tariffs' on and off-peak periods.  A stochastic Ramsey pricing program that maximizes social welfare subject to the revenue break-even constraint for the regulated utility is formulated. Performance of several NEM X policies is evaluated using real and synthetic data to illuminate impacts of NEM policy designs on social welfare, cross-subsidies of prosumers by consumers, and payback time of DER investments that affect long-run DER adoptions.

\end{abstract}
}

\section{Introduction} \label{sec:intro}
Net energy metering (NEM), since its early realization in 1979\footnote{\url{https://en.wikipedia.org/wiki/Net_metering}}, has been one of the major driving forces for the phenomenal growth of the behind-the-meter (BTM) distributed energy resources (DER) such as distributed solar from rooftop photovoltaic (PV). NEM is a class of policies under which a customer with BTM generation resources
is charged for its {\em net-consumption} and credited for the {\em net-production} injected into the grid. Herein, a customer capable of BTM generation is called a {\em prosumer} and one without such capabilities a {\em consumer.}

  Early NEM policies, generally referred to as NEM 1.0,  sets the rate of net-consumption in \$/kWh  (a.k.a. {\it retail rate})  equal to the rate of compensation  (a.k.a. {\it sell rate}) for net energy injection to the distribution grid.  On the one hand, NEM 1.0 provides strong incentives for  adopting BTM
  DER such as solar and storage.   On the other hand, NEM~1.0  creates revenue shortfalls for the distribution utility, equivalently forcing the utility to purchase electricity from prosumers at a retail price significantly higher than the wholesale price \cite{NBERw24756}. As DER adoption grows, a regulated utility may have to increase the retail price to be revenue adequate. Such price increases induce even stronger incentives for additional DER adoptions, which leads to even higher retail price increases.  Theoretically, such a positive feedback loop may result in price instability, leading to a ``death spiral'' scenario \cite{CAI2013830,DARGHOUTH2016713,8675474}.

 Mounting concerns on {\em cross-subsidies} have also been raised.   By realizing significant bill savings under NEM 1.0,  prosumers effectively shift
 a part of their obligation for the grid operation costs to consumers without DER \cite{30415}. Such a cross-subsidy of prosumers by consumers raises normative questions of fairness \cite{Borenstein_solarInequity,PICCIARIELLO201523}.

There have been considerable recent changes to NEM 1.0 by state regulators to address issues of  revenue adequacy and cost allocation among customer groups.  Most notable are the set of new NEM policies currently implemented in California \cite{CPUC_NEM3}, Arizona \cite{nem_state_review_2020}, and New York \cite{CBC_NYstate},  that differentiate the retail and sell rates and/or impose some form of fixed charges such as grid connection or capacity-based charges (CBC). Broadly referred to as NEM 2.0, these policy variations   affect prosumers' consumption choices, consumers' DER adoption decisions, and social welfare distribution that includes prosumer/consumer benefits and societal benefits from decarbonization.  Successor policies of NEM 2.0 (NEM 3.0) are being considered and actively debated  among regulators and industry stakeholders \cite{CPUC_NEM3}.

With evolving NEM policies and the lack of  analysis and modeling  tools, it is challenging to delineate the impacts of the various policy choices within the NEM tariff families.  This paper aims to develop a unified treatment of the NEM policies under an inclusive tariff model and provide analytical and empirical characterizations of prosumers' consumption behavior and impacts of NEM policies on social welfare, cross-subsidies, and DER adoptions.

%https://documents.acer.europa.eu/Official_documents/Position_Papers/Position%20papers/WP%20ACER%2001%2017.pdf

\subsection{Related work}
Extensive literature exists on NEM 1.0 on engineering and economic aspects. The welfare implications and subsidies have been analyzed and empirically evaluated in \cite{8194919,CLASTRES2019100925,Burger2019}, focusing on the efficiency of two-part tariffs with  uniform connection charges.
The adoption dynamics of BTM DER technologies under NEM 1.0 is investigated in \cite{LAWS2017627,DARGHOUTH20115243}, and the stability of the DER adoption processes, the potential of death spiral phenomenon, and mitigation effects of connection charges are considered in \cite{8675474,CAI2013830,DARGHOUTH2016713,LAWS2017627}.  The cross-subsidy issue under NEM 1.0 is considered in \cite{PICCIARIELLO201523, 8194919}.

There is also growing literature on various aspects of NEM 2.0, although a systematic study is lacking. The authors in \cite{8400476} analyze the impact of community solar under three different metering and billing arrangements, including some features of NEM 2.0. Their results show that NEM 1.0 and NEM 2.0 are favorable for community solar customers.  Numerical studies of the economic feasibility of solar plus storage packages under different rate designs, including the sub-retail sell rate feature of  NEM 2.0 are presented in \cite{31935}.  Studies showed that the implemented NEM 2.0 in California yields exaggerated prosumer bill-savings that are higher than the costs of service, which leads to a lager gap between the retail rates and the marginal cost of electricity as the DER adoption evolves \cite{CPUC_NEM3, Next10Report}.

Beyond NEM 2.0, the recent NEM 3.0 rulemaking in California \cite{CPUC_NEM3} presented several alternative rate structures, including adopting connection charges, and CBC to reduce
volumetric-based cost recovery, and highly time-differentiated rates to improve the aggregate load factor. Value of DER (VDER) compensation is also proposed in \cite{Next10Report} to improve equity and cost-reflectivity.

Outside the U.S. markets, there is a global presence of the NEM policies \cite{NEMinAisa:20IEEEaccess,RAMIREZ_NEMEurope:17EnergyPolicy,SHAWWILLIAMS_NEMaustralia:20EnergyPolicy}.
Not captured by the NEM tariff family is the feed-in tariff (FiT) popular in EU, UK, and Japan \cite{RAMIREZ_NEMEurope:17EnergyPolicy}.  Indeed, an EU energy regulator white paper \cite{EUreport:17}  recommends NEM to be avoided to ensure proper cost allocation.  FiT prohibits self-consumption of PV production and enables the regulator to incentivise PV adoption and allocate the costs of network use  by setting separately the retail and sell rates. Although FiT  is not a NEM policy, the analysis developed in this work can be applied directly to the problem under FiT. See Sec. \ref{sec:discussion} for a discussion.

%https://documents.acer.europa.eu/Official_documents/Position_Papers/Position%20papers/WP%20ACER%2001%2017.pdf

\subsection{Summary of results and contributions}
The main contribution of this work is fourfold.  First, we formulate an inclusive NEM X policy model that captures features of existing NEM policies that have been implemented or are being considered for future policy reforms.  The NEM X model provides an analytical framework to evaluate and compare different implementations of NEM policies, complementing some of the existing empirical studies  \cite{Burger2019,CLASTRES2019100925,DARGHOUTH2016713,8675474,8194919}.  The NEM X model parameters  include retail and sell rates, fixed charges, and  dynamic (e.g., TOU) pricing parameters.

Second, we obtain an analytical characterization of the optimal prosumer consumption decision under the NEM X policy model with differentiated retail and sell rates.  Specifically, we show in Theorem~\ref{thm:structure} that the optimal consumption policy for a prosumer has a two-threshold structure.  When the BTM DER production is below a predetermined lower threshold, it is optimal for the prosumer to net-consume.   When the DER production is above a predetermined  upper threshold, it is optimal for the prosumer to net-produce.
When the DER production is between the two thresholds, it is optimal for the prosumer to be a net-zero energy consumer by matching its consumption with its local DER production. Closed-form expressions of the consumption levels and production thresholds are obtained.  The special structure of the prosumer's  optimal consumption policy is especially significant in the analysis of the regulator's rate-setting decision process.

Third, we consider the rate-setting decision process of the regulator under NEM X, taking into account the stochasticity of the BTM DER. In particular, we formulate a {\em stochastic  Ramsey pricing problem} where the parameters of NEM X are chosen to maximize overall social welfare subject to the revenue adequacy constraint that ensures regulated utility cost recovery.  To this end, social welfare is defined as the sum of consumer/prosumer surpluses (private benefits) and total environmental benefits (external benefits) \cite{JianjunExternality:20SusMDPI}.

Finally, we evaluate the impacts of NEM X parameters on social welfare, cost-shifts from prosumers to consumers,  and  payback time of DER investments.  Using real and synthetic data to compare different implementations of NEM-X policies,  we demonstrate, in a short-run analysis, that the transition from the Californian version of NEM 1.0 to NEM 2.0 reduces cross-subsidies and increases social welfare.  However, NEM 2.0 results in a longer payback time for prosumers' DER investments, which may decelerate DER adoption in the long run. Our results elucidate the tensions and tradeoffs among maximizing social welfare, reducing cross-subsidies, and shortening payback time, which affects the long-term growth of DERs.

Our analysis and numerical results help in gaining insights into the effectiveness of various policy realizations, confirming several recent conclusions by economists in their empirical studies. For instance, our analytical characterization of  prosumers' decisions shows that, under NEM 2.0 with differentiated retail and sell rates, prosumers are incentivized
toward more self-consumption rather than bill saving, which is consistent with the conclusion in \cite{Borenstein_canNetMetering,31935}. Our empirical results in Sec.~\ref{sec:simulation} demonstrate that sell-rate reductions bring the retail rate closer to the social marginal cost (SMC) rate, reducing the gap between prosumers and consumers surpluses with increased social welfare, as suggested in \cite{Next10Report}. We also illustrate how an SMC-based sell-rate can effectively mitigate cost-shifts from prosumers, albeit at the cost of prolonged BTM DER payback time, which potentially stalls DER adoption.  Our  numerical results provide evidence that NEM 2.0 and its successors can   achieve higher levels of social welfare compared to NEM 1.0.

\subsection{Notations and nomenclature}
We use boldface for vectors and matrices.
Key designations of symbols are given in Table \ref{tab:symbols}.

 {\small
\begin{table}[htbp]
\caption{\small Major designated symbols (alphabetically ordered).}\label{tab:symbols}
\begin{center}
\vspace{-2em}
\begin{tabular}{|ll|}
\hline\hline
${\bf 0}, {\bf 1}$: & vector of all zeros and ones.\\
$C$: & utility cost function.\\
$\chi(\cdot)$: & indicator function.\\
$\dbf_n, \dbf$: & consumption bundle in billing cycle $n$.  \\
$\dbf^\pi, \dbf^\pi(r)$: & the optimal consumption bundle.\\
$\dbf^\ast_p, \dbf^\ast_c$: & optimal prosumer and consumer consumption.\\
$d^+_i, d^-_i, d^o_i $: & the optimal consumptions of device $i$.\\
$\Delta P^\pi$: & bill savings due to BTM DER under tariff $\pi$.\\
$\Ec$: &  environmental and health benefits of BTM DER.\\
$\gamma$: & the fraction of the population being prosumers.\\
$P^{\pi}:$&  payment schedule under tariff $\pi$.\\
$\pi=(\pi^+,\pi^-\pi^0)$: & NEM X tuple (buy rate, sell rate, fixed charge). \\
$\pi^\omega, \pi^{smc}, \pi^e$: & wholesale, SMC, and environmental prices.\\
$\psi^\pi_\gamma$:& cost-shifts under tariff $\pi$ and adoption level $\gamma$.\\
$r_n, r$: & behind-the-meter renewable in billing cycle $n$  \\
$d^+, d^-$: & thresholds of the optimal prosumer policy\\
$\mathbb{R}^M,\mathbb{R}_+^M$ & sets of $M$ dim. real and positive real vectors.\\
$S^\pi,S^+, S^-$: & prosumer surpluses under $\pi, \pi^+$, and $\pi^-$.\\
$S^\pi_c, S^\pi_u$: & consumers and utility surpluses under $\pi$.\\
$t_{pb}^\pi$: & DER payback time under tariff $\pi$.\\
$U(\cdot), U_i (\cdot)$: & utility functions\\
$V(\cdot), V_i (\cdot)$: & marginal utility functions\\
$W^\pi_\gamma$:& social welfare under tariff $\pi$ and adoption level $\gamma$.\\
$\xi$: & DER installation cost.\\
$y_n$: & net energy consumption of both customer classes.\\
$z_n, z$: & net energy consumption in billing period $n$\\
\hline\hline
\end{tabular}
\end{center}
\end{table}
}

\section{NEM X Tariff Model} \label{sec:model}
We present an engineering-economic model that captures essential characteristics of various existing and proposed NEM-based tariffs and the underlying prosumer decision processes to be incorporated by the rate-setting regulator.

\subsection{A multi-timescale decision model}
The decision processes in retail markets with active consumer participation involve three timescales. For a household equipped with a smart home energy management system, the decision process is constrained by the sensing periods of the sensors that measure the BTM DER, environmental parameters such as temperature and humidity, and the loads consumption levels. Typically, this is the fastest timescale. In our presentation, this timescale is normalized to one.

For the utility under a NEM policy, the {\em net consumption} is computed and billed within a billing period equal to or greater than the prosumer’s decision period. The billing period used in practice ranges from 15 or 30 minute intervals, or on an hourly basis. The NEM tariff model discussed in Sec II assumes this particular timescale.

For the regulator who has the responsibility to set the retail rate of consumption and sell rate of production, the timescale of the rate-setting process is typically on a monthly or yearly basis. Once the NEM parameters are set, it is fixed until the regulator approves the new rate in the next rate-setting cycle. In evaluating the performance of the NEM policies in Sec \ref{sec:short}, social welfare and other metrics are computed based on this time scale.

Throughout the paper, we refer to the decision process of the household EMS as the {\em prosumer decision process} and the rate-setting process as the {\em regulator's decision process}. The short-run analysis considered in this paper  focuses on a single tariff setting period under which the rates are fixed.

\subsection{NEM X tariff model}
We present the NEM  X tariff model where ``X'' stands for NEM variations (\eg NEM 1.0 and NEM 2.0) that have been implemented in practice or under consideration by the regulators.  See \cite{SMITH2021106901,nem_state_review_2020} for a list of NEM implementations that are versions of the NEM X model presented here.

Let  $z$ be the net-energy consumption in a billing period.  Using the indicator function  $\chi(\cdot)$, $\chi(x) = 1$ if $x\ge 0$ and $\chi(x)=0$ otherwise,  the payment from the prosumer during a billing period is given by
\bea
P^{\pi} (z) &=& (\pi^+ z) \chi(z) + (\pi^- z)(1-\chi(z)) + \pi^0, \label{eq:model}
\eea
where the NEM X parameter tuple $\pi=(\pi^+,\pi^{-}, \pi^0)$ is defined as follows:
\bitem
\item[$\pi^+$:]   the retail rate (a.k.a. buy rate). $\pi^+$ is the per kilowatt rate of the net consumption, when $z>0$.
\item [$\pi^-$:] the sell rate  (a.k.a  compensation rate). $\pi^-$ is the per kilowatt rate of compensation to the prosumer, when $z< 0$.
\item[$\pi^0$:] is the non-volumetric surcharge such as connection (delivery) charges, CBC \cite{CBC_NYstate, CPUC_NEM3} or grid access charges (GAC), and income-based charges \cite{Next10Report}.
\eitem
 Note that the so-called Full NEM (monthly or annual billing period), and net-billing or net purchase and sale (hourly or sub hourly billing period) in \cite{8400476} are special cases of (\ref{eq:model}).

\begin{figure}[htbp]
    \centering
    \includegraphics[scale=0.6]{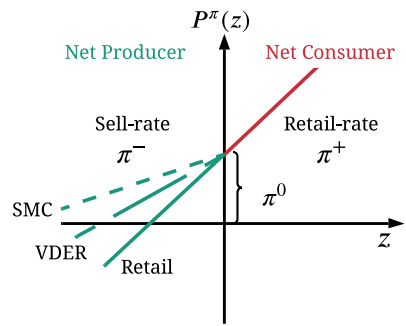}
    \caption{NEM X tariff model in a billing period.}
    \label{fig:depiction_of_consumerPayment}
\end{figure}

The payment expression in (\ref{eq:model}) is further depicted in Fig.\ref{fig:depiction_of_consumerPayment} under different sell-rates, where a net-consumer ($z>0$) faces the retail rate $\pi^+$, and a net-producer faces the sell rate $\pi^-$ ranging usually from the SMC rate tied to the wholesale market to the full retail rate.  The sell rate $\pi^-$ is also referred to as value of DER (VDER), representing the value of the exported energy from prosumers to the utility.

In its full generality, NEM X tariff can be a dynamic policy, depending on the time of use.  Specifically, let $\pi_n = (\pi^+_n, \pi^-_n, \pi_n^0)$ be the NEM-X parameter in the net billing period $n$.  The payment in the $n$th billing period is given by (\ref{eq:model}) with $\pi_n$ and net-consumption $z_n$.

\subsection{NEM 1.0, NEM 2.0, and potential successors}
\subsubsection{NEM 1.0} The conventional net-metering tariff (NEM 1.0) has the same retail and sell rates\footnote{The inclined block rate (IBR) requires a slight generalization where the payment schedule involves multiple affine segments.}.
With $\pi^-=\pi^+$, the payment schedule of NEM 1.0 is given by
\[
P^{\rm\tiny \pi_{1.0}} (z) = \pi^+ z + \pi^0,
\]
where the connection charge $\pi^0$ is in the range of a few to tens of dollars.  The volumetric charge $\pi^+$ is usually considerably higher than the wholesale rate of electricity \cite{Next10Report}.

\subsubsection{NEM 2.0}  NEM 2.0's feature of differentiated retail and sell rates is captured directly in the NEM X model.
The price differential $\delta=\pi^+-\pi^-$ ranges from a few cents/kWh to a significant level, when $\pi^-$  approaches to the SMC rate.

Another feature of some implementations of NEM 2.0 is the TOU pricing involving on and off peak rates, $\pi_{\mbox{\tiny on}}=(\pi_{\mbox{\tiny on}}^+, \pi_{\mbox{\tiny on}}^-, \pi^0)$ and
 $\pi_{\mbox{\tiny off}}=(\pi_{\mbox{\tiny off}}^+, \pi_{\mbox{\tiny off}}^-, \pi^0)$, respectively \cite{CPUC_NEM3}.

\subsubsection{NEM 3.0 and beyond}  NEM 2.0 can be seen as a set of transitional policies from NEM 1.0 to the more cost-reflective tariff policies, broadly referred to as NEM 3.0. Four major potential directions of NEM 3.0 and successor policies are under discussion. One is to introduce discriminative fixed charges to prosumers based on their DER capacities or income levels \cite{CBC_NYstate,CPUC_NEM3}. Another is to discriminate in volumetric charges by treating prosumers as a distinct customer class \cite{volumetric_discrimination}. Furthermore, additional sell-rate reduction is expected to be prevalent in NEM 3.0. In some states  \cite{nem_state_review_2020,CPUC_NEM3}, the sell-rate has already been discounted to a level close to the SMC rate ($\pi^-=\pi^{\mbox{\tiny SMC}}$). Lastly, policy reforms can come in the form of shortening the billing period, which prolongs the payback time of DER \cite{CPUC_NEM3}.

\section{Optimal Prosumer Decision under NEM X} \label{sec:prosumer}
Under NEM X, the prosumer decision process is to optimize its consumption to meet its lifestyle requirements and household scheduling constraints, given the  measured (and forecasted) BTM generation and other sensor data. To avoid burdensome notations, we consider the special case when the prosumer decision process has the same timescale as that of the utility's billing period.

 We begin with the standard microeconomics formulation of the household decision problem involving $M$ consumption devices such as in lighting, heating/cooling, and EV charging.  With measurements of  DER production and under the NEM X tariff, the household energy management system sets the optimal consumption bundle $\dbf \in \{(d_1,\cdots, d_M)| d_i \ge 0\}$ to maximize its utility or surplus subject to a budget constraint. Specifically, we consider
 the following surplus maximization within a fixed billing period:
 \beq
 \begin{array}{lll}
\Pc: &  \underset{\dbf  \in \mathbb{R}^M}{\rm maximize}& S^\pi(\dbf):=U(\dbf) - P^{\pi} (z) \\
&{\rm subject~ to} & z = {\bf 1}^{\T} \dbf -r \\
 & & P^{\pi} (z ) \le B\\
 & & {\bf 0} \le   \dbf  \le \bar{\dbf}
 \end{array}
 \label{eq:prosumerOPT}
 \eeq
 where $U(\cdot) $ is the utility function of the consumption bundle, $r \ge 0$ the (forecasted) BTM renewable generation,  $\pi$ the NEM X parameter, $B$ the budget constraint, and $\bar{\dbf}$ the consumption upper limits.

The prosumer optimization in (\ref{eq:prosumerOPT}) is convex, when the utility function is concave.  Given the specific form of $U$, the solution can be obtained analytically or numerically. Machine learning methods can be used when $U$ is unknown. 

A different angle view of this model is to think of $M$ as a group of
households owning a central DER. This is referred to as community net-metered solar or virtual net-metering.

\subsection{Optimal prosumer decision}
We exploit the piecewise linear property of the NEM X tariff model (\ref{eq:model}) for a structural solution of (\ref{eq:prosumerOPT}) by making a few simplifying assumptions:
\bitem
\item[A1:]  {\em Additivity and concavity.} For a household having $M$ electric devices, the utility of the consumption bundle $\dbf$ is additive
\[
U(\dbf)=\sum_{i=1}^M U_i(d_i),
\]
where $U_i(d_i)$ is the utility function of device $i$.  For all $i$, the utility function $U_i$ is concave, monotonically increasing,  and  continuously differentiable with {\em marginal utility} $V_i(x) = \frac{d}{dx} U_i(x)$.
\item[A2:]  The sell rate is no higher than the retail rate, $\pi^+\ge \pi^-$.
\item[A3:] {\em Non-binding budget constraint.} We assume the budget constraint to be non-binding.
\eitem
 A1-A2 are mostly standard in economic analysis.  The retail rate is almost always no lower than the sell rate.
The non-binding budget constraint is acceptable when the overall electric bill is a small fraction of the disposable income of the household.

The following theorem gives a closed-form characterization of the optimal prosumer decision and a simple computation procedure to schedule consumptions.

\begin{theorem}[Prosumer decision under NEM X]  \label{thm:structure}
Given NEM parameter $\pi=(\pi^+,\pi^-,\pi^0)$ and the marginal utilities $(V_1, \cdots, V_M)$ of consumption devices, under A1-A3 and non-degeneracy condition of (\ref{eq:prosumerOPT}), the optimal prosumer consumption policy is given by two thresholds
\beq
\begin{array}{l}
d^+:=\sum_i  \max \{0,\min\{V_i^{-1}(\pi^+),\bar{d}_i\}\},\\
d^-:=\sum_i \max  \{0,\min\{V_i^{-1}(\pi^-),\bar{d}_i\}\} \ge d^+\\
\end{array}
\eeq
that partition the range of DER production into three zones:
\ben
\item \underline{Net consumption zone:  $r<d^+$}. The prosumer is a net-consumer with consumption
\beq \label{eq:d_i^+}
d^+_i =\max \{0,\min\{V_i^{-1}(\pi^+),\bar{d}_i\}\} \ge 0,~~\forall i.
\eeq
\item  \underline{Net production zone:  $r>d^-$}. The prosumer is a net-producer with consumption
\beq \label{eq:d_i^-}
d^-_i=\max  \{0,\min\{V_i^{-1}(\pi^-),\bar{d}_i\}\} \ge d_i^+,~~\forall i.
\eeq
\item  \underline{Net-zero energy zone:  $d^+ \le r \le d^-$}. The prosumer is a net-zero consumer with consumption:
\beq \label{eq:d_i^o}
 d^o_i(r)  =  \max\{0, \min\{V_i^{-1}(\mu^*(r)),\bar{d}_i\}\} \in [d^+_i, d^-_i], \forall i
\eeq
where $\mu^*(r) \in [\pi^-, \pi^+]$ is a solution of
\beq
\sum_{i=1}^M \max\{0, \min\{V_i^{-1}(\mu),\bar{d}_i\}\} = r.
\label{eq:r}
\eeq
Furthermore, $d_i^o(\cdot)$ is continuous and monotonically increasing in $[d^+_i,d^-_i]$.
\een
\end{theorem}

{\em Proof:} See Appendix.

\subsection{Structural properties of prosumer decisions}
The structure of the optimal consumption policy in Theorem~\ref{thm:structure} plays a significant role in our analysis.  Here we highlight some of its key properties.

\subsubsection{Two-threshold decision policy and piece-wise linear demand}  
\begin{center}
\begin{figure}[htbp]
\begin{psfrags}
\psfrag{F}[c]{\tiny $S$}
\psfrag{F1}[c]{\tiny $S^-$}
\psfrag{F2}[c]{\tiny $S^+$}
\psfrag{a}[c]{\tiny $d^+$}
\psfrag{b}[c]{\tiny $d^-$}
\psfrag{D}[c]{\tiny  $d$}
\psfrag{Dr}[c]{\tiny  $d^\pi(r)$}
\psfrag{r}[c]{\tiny  $r$}
\psfrag{d}[c]{\tiny  $r$}
\psfrag{g}[c]{\tiny  $d^+$}
\psfrag{f}[c]{\tiny  $d^-$}
\centerline{\scalefig{0.25}\epsfbox{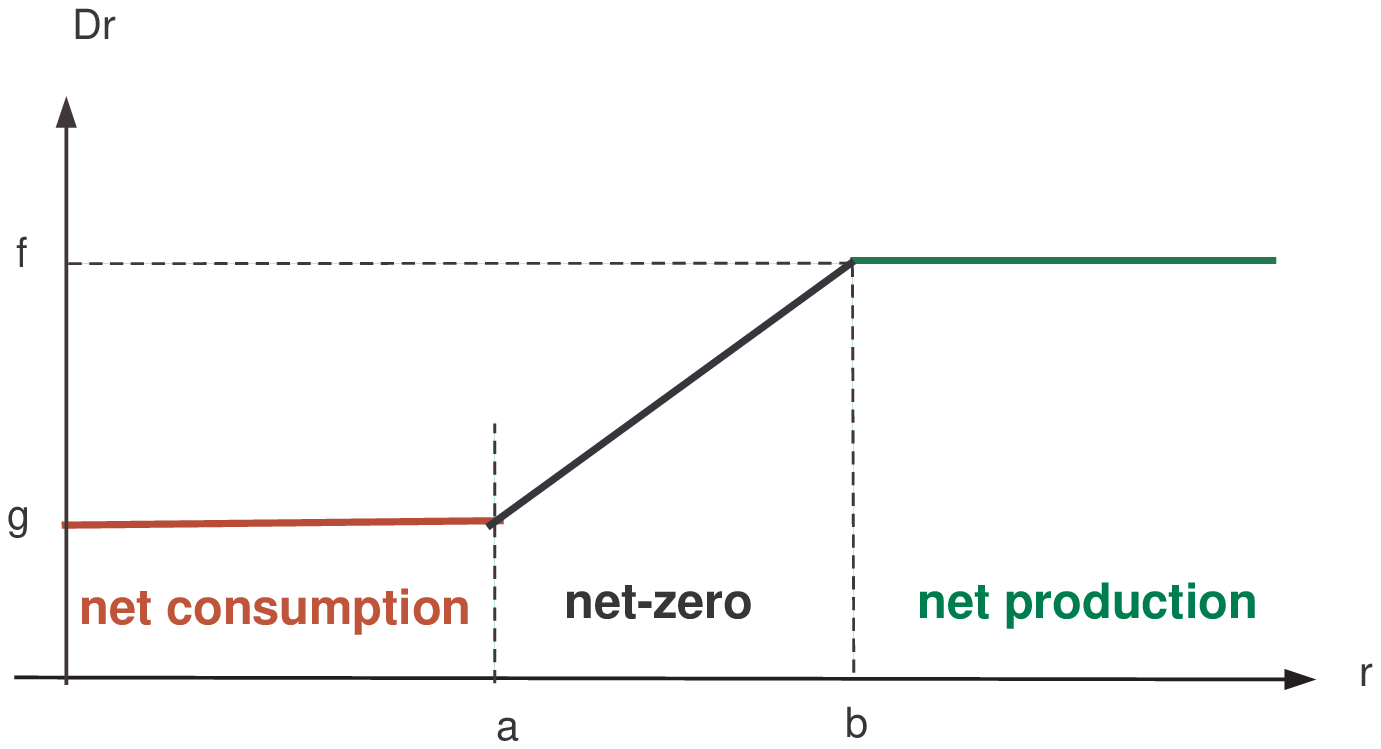}\scalefig{0.25}\epsfbox{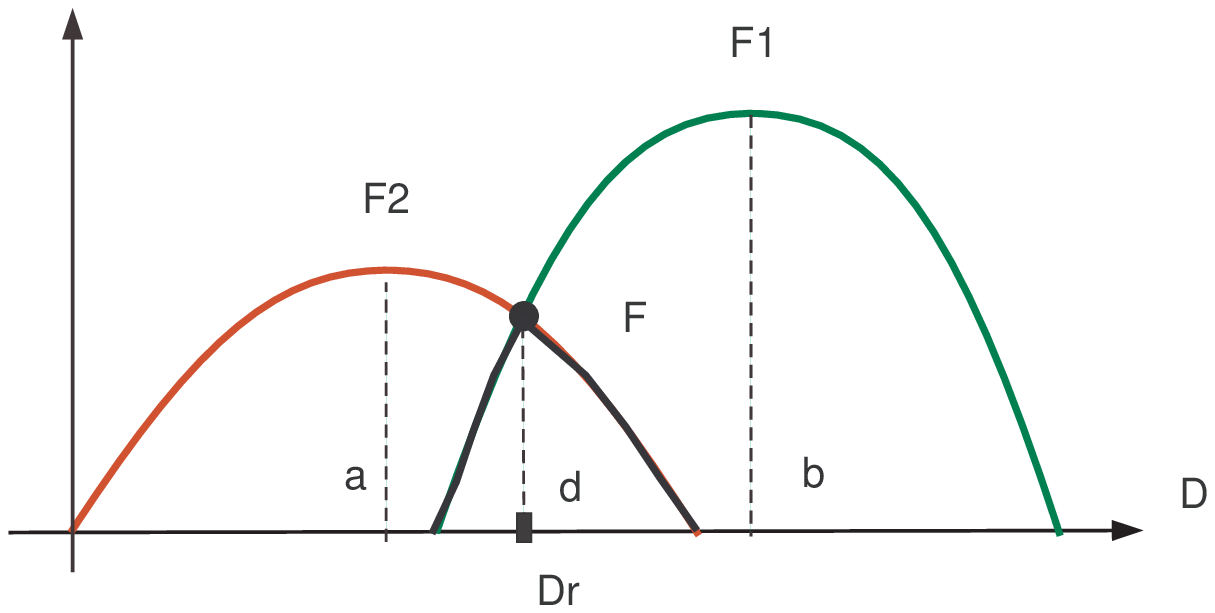}}
\centerline{\scalefig{0.25}\epsfbox{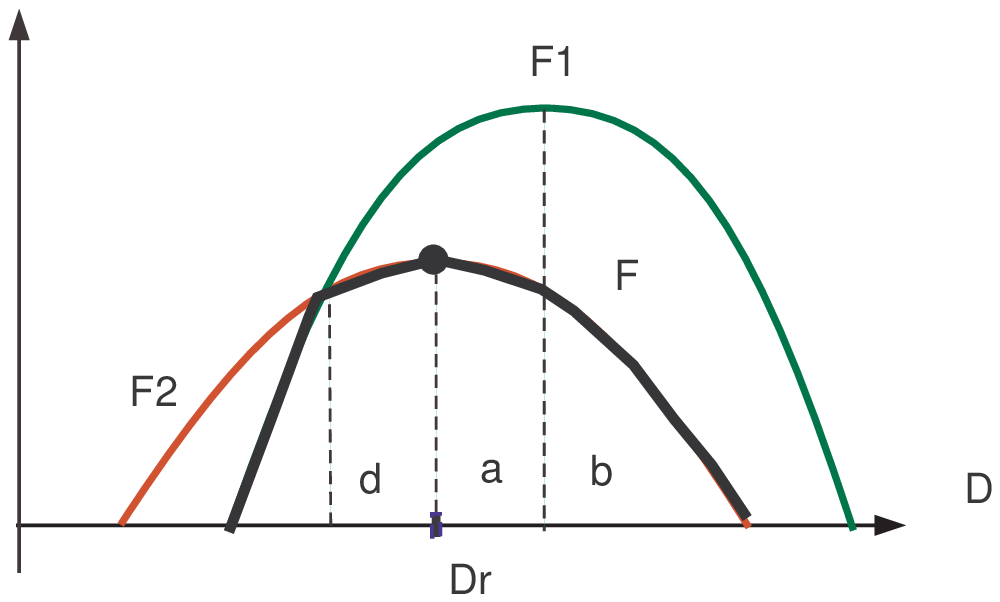}\scalefig{0.25}\epsfbox{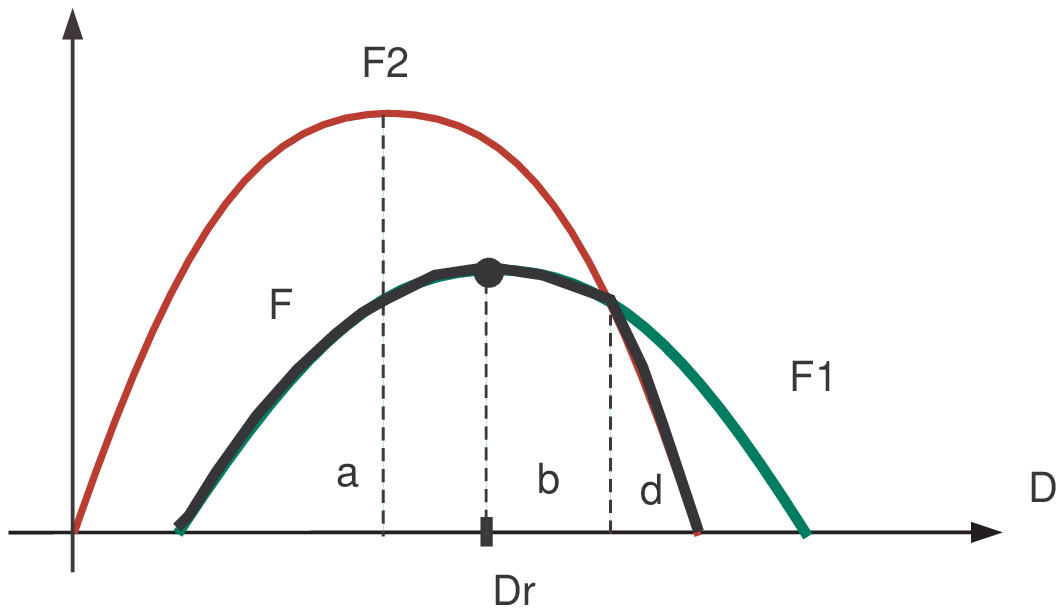}}
\end{psfrags}
\caption{ Structure of optimal prosumer schedule (Note that the figure is not sketched at scale). Top-left: the optimal consumption level for device $i$ as a function of the BTM generation $r$. Rest of  the figures: The surpluses $S^+$ (red) and $S^-$ (green) of the prosumer as functions of demand $d$ under $\pi^+$ and $\pi^-$, respectively.   The surplus function under NEM X is $S$ (black).}
\label{fig:structure}
\end{figure}
\end{center}

Theorem~\ref{thm:structure} shows that the optimal consumption of every device has the same decision zones, which implies that the total demand is a piece-wise linear function of the BTM generation $r$, as illustrated in the top-left panel of Fig.~\ref{fig:structure}. This structure is particularly useful in characterizing statistical properties of prosumer behavior.  Whereas the threshold decision structure is intuitive and  appealing, it is not obvious that the optimal consumption should have a net-zero energy mode over a range of renewables, $d^+ \le r \le d^-$, in which the prosumer's  consumption tracks the renewable.  

The insight into the optimal schedule in
 the net-zero zone ($d^+ \le r \le d^-$), assuming aggregate  consumption of all devices, is shown in the top right panel of Fig.~\ref{fig:structure}, which shows the  prosumer surpluses $S^+(d)$, and $S^{-}(d)$  under $\pi^+$ and $\pi^-$, respectively defined as
\[
S^+(d) :=U(d)-\pi^+(d-r),~S^-(d) :=U(d)-\pi^-(d-r),
\]
with $S^+$ (red) being the surplus when the prosumer consumes and $S^-$ (green) when the prosumer produces.
Note that $S^+$ and $S^-$ achieve their maxima at $d^+$ and $d^-$, respectively, and the two curves  intersect only at $d=r$.

Intuitively, when the renewable generation is abundant $r \geq d^+$, 
the prosumer's increase in the aggregate consumption is a manifestation of the higher benefit of self-consumption (valued at $\pi^+)$ compared to energy net-exportation (valued at $\pi^{-} \le \pi^+$).

\subsubsection{Priority rule for consumption allocation}
\begin{figure}[htbp]
    \centering
    \includegraphics[scale=0.6]{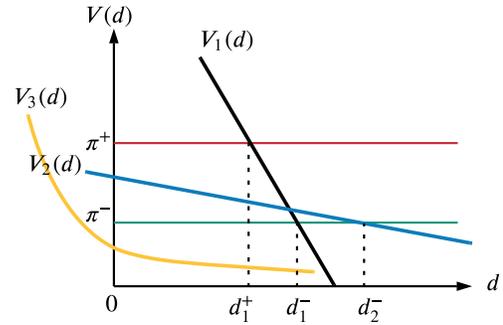}
    \caption{Marginal utilities of three devices.}
    \label{fig:marginalU}
\end{figure}

The optimal schedule given in Theorem~\ref{thm:structure} has immediate microeconomics interpretations based on the role of marginal utility functions as illustrated in Fig.~\ref{fig:marginalU}.

Theorem~\ref{thm:structure} shows that the optimal schedule sets the consumption levels such that all devices have the same marginal utility. Implicitly, the  optimal schedule $(d_i^\pi(r))$ in (\ref{eq:d_i^+}-\ref{eq:d_i^o}) realizes a  priority ranking of consumptions based on their marginal utilities: a device with higher utility of consumption is scheduled to consume more. Specifically, under $\pi^+$, the consumption levels are in the reverse order of their  utilities:  $d_1^+> d_2^+ > d_3^+ = 0$. Under $\pi^-$, $d_2^->d_1^->d_3^-$, which is analogues to the economic dispatch problem of generators in power system operations  where a generator with lower cost is prioritized over those with higher cost.

 The priority rule also determines whether the device is consumed in all three zones or not. Some devices are only consumed when the prosumer has abundant local generation.  Using only the marginal utility at the origin, the optimal consumption pattern has three cases, as illustrated in Fig.\ref{fig:marginalU}. Those devices whose marginal utilities at the origin  are greater than $\pi^+$ will always  be scheduled to consume.   Those devices with marginal utilities at the origin below $\pi^-$ will never be used. Those whose marginal utilities at the original are between $\pi^-$ and $\pi^+$ will only be used when the BTM DER is beyond the net-consumption zone.  Proposition~\ref{prop:loadranking}  formalizes the above intuition.  

 \begin{proposition}[Load priority ranking rule] \label{prop:loadranking}
Given the NEM X parameter $\pi = \left(\pi^+,\pi^-, \pi^0 \right)$, the marginal utilities $(V_1,\cdots,V_M)$, the optimal consumption pattern of every device $i$ is given by
\begin{enumerate}
    \item If $V_i(0) > \pi^+ \geq \pi^-$, device $i$ is always scheduled to consume in one of the three consumption zones.
    \item If $\pi^+ \geq V_i(0) \geq \pi^-$, device $i$ consumes only when the BTM DER is beyond the net-consumption zone.
   \item If $\pi^+ \geq \pi^- \geq  V_i(0)$, it is optimal not to schedule device $i$ to consume.
\end{enumerate}
\end{proposition}
{\em Proof:} See Appendix.

\subsection{Comparative static of NEM X}
We now consider impacts of exogenous parameters such as the level of renewables and NEM X parameter $\pi$ on the endogenous quantities of consumption, surplus, and prosumer payment. The following theorem formalizes the comparative static of the optimal prosumer schedule under NEM X.

\begin{theorem}[NEM X comparative static] \label{thm:comparative}
The optimal consumption level $(d_i^\pi(\cdot))$ and the prosumer surplus $S^\pi(\cdot)$ are monotonically increasing, and the prosumer payment $P^\pi(\cdot)$ is monotonically decreasing.

For every $r$ at the interior of each scheduling zone, as $\pi^+$ increases, the net-consumption zone shrinks, the net-zero zone expands, and the net-production zone stays unchanged.  Meanwhile, for all $i$,
the consumption level $d_i^\pi(r)$  and the prosumer surplus $S^\pi(r)$   decrease monotonically with $\pi^+$.

As $\pi^-$ increases, the net-production zone expands, the net-zero zone shrinks, and the net-consumption zone stays unchanged.  Meanwhile, for all $i$, the net consumption level $d_i^\pi(r)$ and total payment decreases monotonically while prosumer surplus $S^\pi(r)$  increases monotonically with $\pi^-$.

As $\pi^0$ increases, the total payment $P^\pi(r)$ increases, the prosumer surplus $S^\pi(r)$ decreases, and the prosumer consumption $d_i^\pi(r)$ stays unchanged.
\end{theorem}
{\em Proof:} See Appendix.

\begin{table}[htbp]
\caption{Comparative static analysis}
\vspace{-0.5em}
\begin{psfrags}
\psfrag{R}[c]{$r\uparrow$}
\psfrag{P1}[c]{$\pi^+\uparrow$}
\psfrag{P2}[c]{$\pi^-\uparrow$}
\psfrag{P3}[c]{$\pi^0\uparrow$}
\psfrag{z+}[c]{$\small \Rmsc^+$}
\psfrag{z-}[c]{$\Zmsc^-$}
\psfrag{z0}[c]{$\Zmsc^o$}
\psfrag{D1}[c]{$d_i^\pi(r)$}
\psfrag{D2}[c]{$S^{\pi}(r)$}
\psfrag{D3}[c]{$P^\pi(r)$}
\psfrag{D4}[c]{$P^\pi(r)$}
\centerline{\scalefig{0.4}\epsfbox{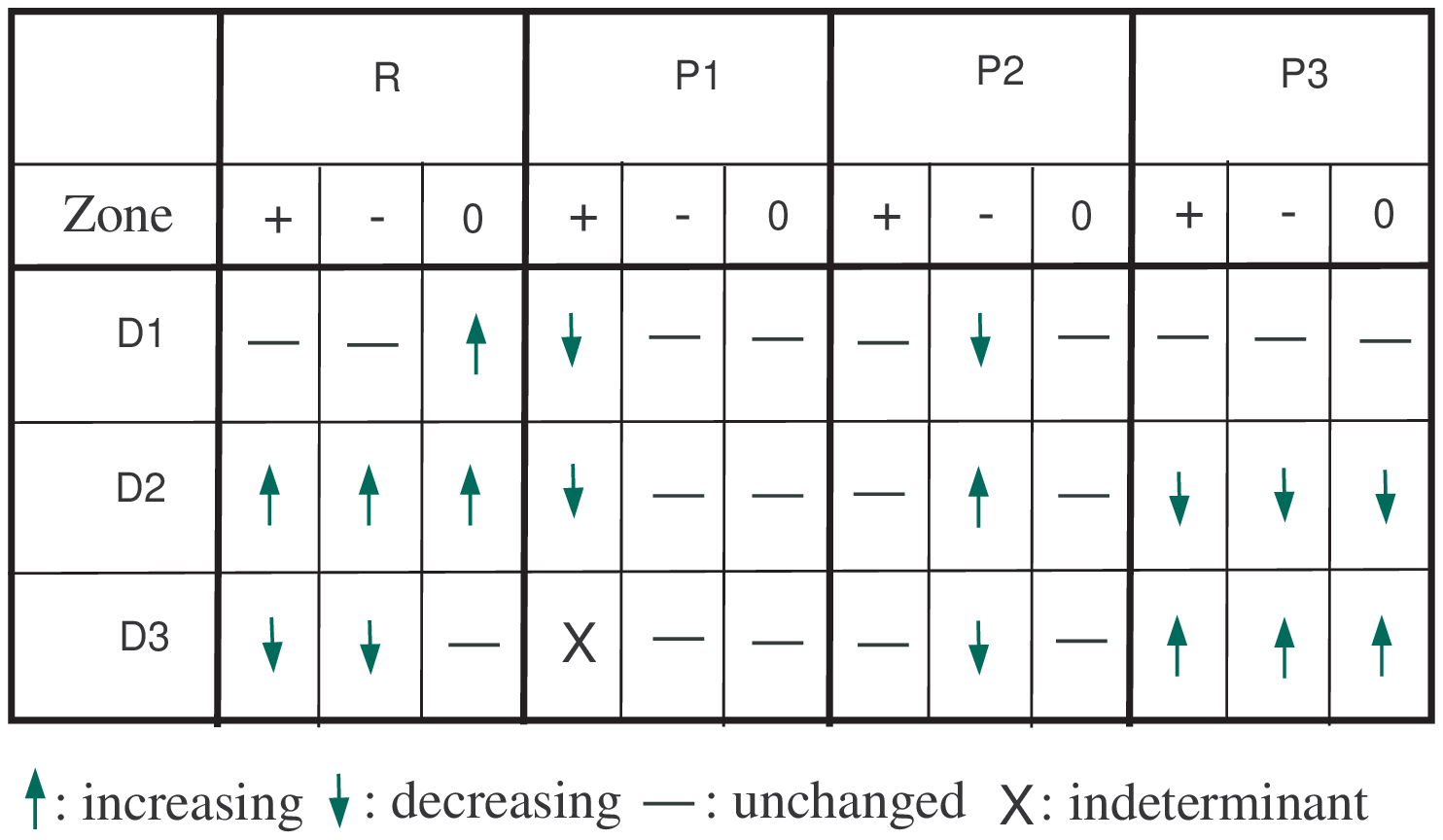}}
\end{psfrags}
\label{tab:comparative}
\end{table}

Table~\ref{tab:comparative} summarizes the detailed comparative static analysis by considering $\epsilon$-increases of the exogenous parameters and examining changes of endogenous quantities at interior points of net-consumption (+), net-production (-), and net-zero energy (0) zones.

Several observations are in order.  First,  under NEM X, prosumer consumption and surplus are monotonic in all scheduling zones with respect to DER generation $r$ and NEM X price parameters.  Prosumer payment is also monotonic except in the net-consumption zone, where $\frac{\partial}{\partial \pi^+}P^{\pi}(r)$ may be positive or negative depending on the value of $r$ and the characteristic of the utility function $U_i(\cdot)$.

Second,  Table~\ref{tab:comparative} gives  additional insights into how a prosumer exploits behind-the-meter generations. In the net-zero energy zone, the consumer increases its consumption with $r$  (thus its utility) without having to pay to the utility company for its consumption.  In both the consumption and production zones, the consumer holds the consumption level constant while minimizing its payments to the utility.   In all three operation zones, the prosumer minimizes its payment to the utility, resulting in cost-shifts to other consumers and prosumers.  See more discussions in Sec.~\ref{sec:short}.

Third,  as $\pi^-$ increases to $\pi^+$, NEM X becomes NEM 1.0.   The production zone enlarges and eventually connects with the consumption zone.  Because it is more profitable to sell extra power to the utility,  the consumer surplus increases, the payment to the utility decreases (more precisely, the payment from the utility increases), and consumption level drops.

\section{NEM X rate setting and short-run analysis} \label{sec:short}
In a retail market where the utility is a regulated monopoly,  the regulator sets the price of electricity subject to the constraint that the utility recovers its costs. To this end, the standard approach is the Ramsey pricing where social welfare is maximized subject to the break-even constraint \cite{brown_sibley_1986}.

\subsection{Stochastic Ramsey Pricing under NEM X}
We assume that the consumer population consists of prosumers (BTM DER adopters) and consumers (non-adopters).
We make a standard assumption that, within each consumer group,  the aggregated demand  can be obtained by a representative customer (prosumer and consumer, respectively).  Such an assumption is valid when consumer/prosumer utilities have the Gorman form \cite[p. 119]{Colell&Whinston&Green:book} and is reasonable for typical household devices.  Under this assumption, the optimal consumption rule given by Theorem~\ref{thm:structure} can be used. The optimal consumption policy, however, is derived based on the specific DER realization. It is, therefore, necessary to develop a stochastic Ramsey pricing formulation.

Ramsey pricing maximizes social welfare.  Suppose that each rate-setting cycle consists of $N$ billing periods for which NEM X parameters are fixed. Let $r_n$ be the random BTM DER production in the $n$th period.  Under NEM X with parameter $\pi$, we define the social welfare $W^{\pi}_\gamma$  as the sum of the expected customers (consumer and prosumer) surplus $S^\pi_c$,  the expected utility surplus $S_u^\pi$, and the expected environmental and health benefits $\Ec(r_n)$ brought by the BTM DER
\beq
W^{\pi}_\gamma= \sum_{n=1}^N \mbbE \bigg( S_c^\pi (\dbf^*(r_n),\gamma) + S^\pi_u (\dbf^*(r_n),\gamma) + \gamma \Ec(r_n)\bigg),
\label{eq:welfare}
\eeq
where $\gamma \in [0,1]$ is the fraction of prosumers in the customer population, and $\dbf^*(r_n)$ is the optimal population consumption under $\pi$.

Given $\gamma$, the expected customer surplus in the billing period $n$ is given by
\beq
\mbbE(S_c^\pi (\dbf(r_n),\gamma)) = \gamma \mbbE(S^\pi(\dbf^*_p(r_n))) + (1-\gamma)S^\pi(\dbf^*_c),
\label{eq:Sc}
\eeq
where $\dbf^*_p(r_n)$ is the optimal prosumer consumption given by Theorem~\ref{thm:structure} and $\dbf^*_c$ the optimal consumption of consumers without DER, also given by Theorem~\ref{thm:structure} with $r_n=0$.

Likewise, the expected  utility surplus in the billing period $n$ is given by
\bea
\label{eq:Su}
\mbbE(S_u^\pi (\dbf^*(r_n),\gamma)) &=&   \gamma \mbbE(P^\pi({\bf 1}^{\T}\dbf^*_p(r_n)-r_n)\\& & + (1-\gamma) P^\pi({\bf 1}^{\T}\dbf^*_c)
   - C(y_n(r_n,\gamma)))\nn,
\eea
where $P^\pi$ is the payment of prosumers in billing period $n$ defined in (\ref{eq:model}), and $C(\cdot)$  is the utility's cost function to meet the total customers' net demand $y_n$ in billing period $n$, defined by
\[
y_n(r_n,\gamma) := \gamma ({\bf 1}^{\T}\dbf_p^*(r_n)-r_n)+ (1-\gamma) {\bf 1}^{\T}\dbf^*_c.
\]
The expected  environmental and health benefit,  modeled as a linear function of total BTM generation $r_n$, is given by
\beq
\mbbE(\Ec(r_n))=  \pi^e  \mbbE(r_n)
\label{eq:Environmental}
\eeq
where $\pi^e$ is the ``price'' by which  DER is valued toward realizing clean energy goals \cite{61024}.

By setting the NEM X parameters such that the utility recovers it operating costs, the Ramsey pricing  is given by the following stochastic social welfare  maximization:
  \beq
 \begin{array}{ll}
\underset{\boldsymbol{\pi} }{\rm maximize}& \sum_{n=1}^N \mbbE\bigg(S_c^\pi (\dbf^*(r_n),\gamma)+ \gamma \Ec(r_n)\bigg)\\
\mbox{subject to} & \sum_{n=1}^N \mbbE(S_u^\pi (\dbf^\ast(r_n),\gamma)) =0.
\label{eq:Ramseymaximization}
 \end{array}
\eeq
The special structure of the optimal consumption given in Theorem~\ref{thm:structure} makes solving the above optimization tractable when
the underlying probability distribution of  $r_n$ is known.   In practice, the above stochastic Ramsey pricing program can be replaced by a scenario-based optimization with representative scenarios of DER productions. Note that not all the NEM X parameters are optimized in practice.

\subsection{Performance of NEM X}
\subsubsection{Social Welfare}  Given NEM X parameter $\pi$, the social welfare of NEM X  is given by $W_\gamma^\pi$ in (\ref{eq:welfare}-\ref{eq:Environmental}).
\subsubsection{Cost Allocation} The retail tariff allows the utility to recover its cost from customer payments. BTM DER reduces the payments of prosumers, thus shifting costs to consumers.  Here we measure cost allocation of NEM X by the \textit{bill-saving} of a prosumer in the billing period $n$ and cost-shift.

\par The bill-saving of a prosumer in the billing period $n$ is defined by the monetary gain as a result of the prosumer's on-site generation, and is given by \cite{EdisonReport_Borlick_Wood}
\beq
    \Delta P^\pi (r_n):= P^\pi({\bf 1}^{\T}\dbf^*_c) - P^\pi({\bf 1}^{\T}\dbf^*_p(r_n) - r_n).
\eeq
Normally $\Delta P^\pi (r) \geq 0$. When fixed charges such as CBC are involved, the bill-saving of a prosumer may be negative. Note that under the special case of NEM 1.0, the volumetric part of bill savings is simply $\pi^+ r_n$.

This cost shift represents a form of cross-subsidy of prosumers by consumers.   A measure of such cross-subsidy is the {\em expected cost-shift} $\psi^\pi_\gamma $  defined by the difference between the expected prosumer bill-saving and the utility avoided cost (for not having to procure power) because of DER generation:
\begin{equation}
    \label{eq:ProfitDefficit}
    \psi^\pi_\gamma = \sum_n \gamma  \mathbb{E}\left(\Delta P^\pi\left(r_n\right)- \pi^{\mbox{\tiny SMC}} r_n \right),
\end{equation}
where $\pi^{\mbox{\tiny SMC}}$ is the social marginal cost price of electricity \cite{EdisonReport_Borlick_Wood,Next10Report}.

\subsubsection{DER market potential}  The market potential is a short-run measure of a product diffusion in a market.  The market potential of DER is primarily determined by the payback time of the DER investment from the achieved bill-savings under a particular NEM X policy \cite{8675474,beck}.  Here we use the net present value (NPV) method with discounted bill-saving for the expected payback time:
\beq
\label{eq:paybacktime}
    t_{pb}^\pi(r,\xi)= \frac{\xi}{\mathbb{E}\left(\Delta P^\pi(r)\right)},
\eeq
where $\mathbb{E}\left(\Delta P^\pi(r)\right) > 0$, and $\xi$ is the DER installation cost, which may include any state rebates or investment tax credits. To account for the time value of money in the payback calculation, we utilize the time-to-net payback time expression in \cite{drury_denholm_margolis_2011}, as the following:
\begin{equation}
\label{eq:TNPpaybackTime}
    t_{pb}^\pi(r,\xi) =\min_{t^\ast}\left\{t^\ast: \sum_{t=0}^{t^\ast} \left(\frac{1-\nu}{1+\zeta}\right)^t  \mathbb{E}\left(\Delta P_t^\pi(r_t)\right) \geq \xi \right\},
\end{equation}
where $\nu,\zeta \in [0,1)$ are the BTM DER system degradation factor and interest rate, respectively, $\alpha$ is the compounding rate.

The payback time expressions greatly depend on the expected bill savings. Under NEM X, decreasing $\pi^-$ decreases bill savings, which prolongs the payback time. The shortest payback time is therefore achieved  under NEM 1.0, where the sell-rate is at its highest level $\pi^-= \pi^+$. Under uniform two-part tariffs, in addition to the role of $\pi^-$, the lump sum connection charge reduces the breakeven retail rate in (\ref{eq:Ramseymaximization}), which in turn prolongs the expected payback time of the BTM DER \cite{8675474}. A discriminatory two-part tariff, on the other hand, imposes a twofold effect on prosumers: 1) by extracting more of the prosumer surplus through the additional fixed fee, 2) by reducing the breakeven retail rate, which extends the payback period. Therefore, interestingly, both uniform and discriminatory two-part tariffs affect the prosumer surplus more than the consumer surplus, but the latter enables the regulator in designing a tariff that achieves a specific targeted and DER adoption sustaining payback time period without impacting consumers.

The expected payback time is a key factor in a consumer's DER adoption decision; it has a direct impact on the trajectory of long run DER adoption.

\section{Numerical Results}\label{sec:simulation}
To estimate and analyze the different NEM X tariff models and the role of utility rate structure in practical settings, a hypothetical distribution utility facing the wholesale price\footnote{The day-ahead LMP data is taken from CAISO SP15 for the period June-August, 2019. The data can be found at: \url{http://oasis.caiso.com/mrioasis/logon.do}.} $\pi^\omega$ in California was assumed. The NEM X parameter was periodically set based on Sec.~\ref{sec:short}, which incorporated the prosumers and consumers optimal consumption decisions obtained in Sec.~\ref{sec:prosumer}.

\begin{table}[htbp]
\centering
\caption{Description of case studies}
\label{tab:Plotcasesarxiv}
\resizebox{0.48\textwidth}{!}{%
\begin{tabular}{@{}ccccccc@{}}
\toprule \toprule
\# & Name        & Tariff   & Rate & \begin{tabular}[c]{@{}c@{}}Compensation \\ (\$/kWh)\end{tabular} & \begin{tabular}[c]{@{}c@{}}Fixed\\  Charges\end{tabular} & Discrimination$^\ddagger$ \\ \midrule
1  & NEM 1.0     & one-part & IBR$^\ast$  & $\pi^- = \pi^+$                                                  & --                                                       & No             \\
2  & NEM 2.0     & one-part & TOU$^{\mathsection}$  & $\pi^- = \pi^+-0.03$                                             & --                                                       & No             \\
3  & NEM SMC & one-part & TOU  & $\pi^- = \pi^{smc}$                                              & --                                                       & No             \\
4  & NEM CBC & two-part & TOU  & $\pi^-  = \pi^+ -0.03$                                           & 10.93$^\dagger$                                                    & Yes            \\ \bottomrule
\bottomrule \end{tabular}}
\begin{adjustwidth}{0pt}{0pt} \footnotesize{  \begin{flushleft} \vspace{0.2cm} $^\ast$ Similar to PG\&E, IBR has two blocks, with a 20\% higher price for above baseline usage.\\
$^{\mathsection}$ TOU parameters for all case studies are similar to PG\&E TOU-B, i.e. 1.5 peak ratio and a 16 -- 21 peak period.\\
$^\dagger$ CBC value is expressed in \$/kWDC PV/month, and taken from \cite{CPUC_NEM3}.\\ $^\ddagger$ We refer to the residential customers inter-class discrimination between prosumers and consumers.
 \end{flushleft}}
\end{adjustwidth}
\end{table}

\subsection{Data sources and settings of numerical studies}
The aggregated demand was obtained by a representative prosumer and consumer model, where the demand of each of the considered typical household devices has a Gorman form utility function. The utility parameters of each household device were estimated from historical retail prices and consumption data.  See Appendix B.  To solve the prosumer problem, we adopted an average 5.1 kWDC/household PV installed capacity and the corresponding hourly production profile of solar PV\footnote{The solar data profile from California solar initiative is a 15-Minute interval PV data for the period June-August, which can be found at: {\url{https://www.californiadgstats.ca.gov/downloads}}.}.
In setting the NEM X parameters, we estimated the utility fixed costs $\theta$ which is part of $C(\cdot)$ in (\ref{eq:Su}) using publicly available revenue, MWh sales, and number of customers data of Pacific Gas and Electric Company (PG\&E)\footnote{Revenue, sales, and number of customers of PG\&E data was taken from EIA over the years from 2016-2019: \url{https://www.eia.gov/electricity/data/state/} .}. The estimated average daily fixed cost to serve a single residential customer was $\theta_{PGE} = \$2.86/day$. The SMC rate was assumed to be the sum of $\pi^\omega$ and the non-market cost of pollution that was reflected on the retail price \cite{Next10Report}\footnote{The non-market cost of pollution was calculated based on the avoided non-energy cost due to BTM DER, which was estimated by \cite{Borenstein_avoidedCost} to be \$0.012/kWh, and the renewable portfolio standard (RPS) compliance benefits estimated to be \$0.018/kWh, as mentioned in: \url{https://www.sce.com/regulatory/tariff-books/rates-pricing-choices/renewable-energy-credit}.}. To compute (\ref{eq:Environmental}), the price $\pi^e$ was quantified at \$0.035/kWh solar from \cite{61024}.

\par We evaluated performance of four NEM X policies with parameters shown in Table.~\ref{tab:Plotcasesarxiv}, aimed at  gaining insights into characteristics of the initial policy NEM 1.0, the current implemented policy NEM 2.0, and two NEM successor policies involving SMC-based sell rate (NEM SMC) and capacity based charge (NEM CBC). In all simulation results, an average PV system cost of $\xi=$ \$4500/kW was used\footnote{The average 2019 solar cost data for systems less than 10kW in California can be found at: \url{https://www.californiadgstats.ca.gov/charts/nem}.}.

% Please add the following required packages to your document preamble:
% \usepackage{booktabs}

\subsection{Social welfare}
\begin{figure}[htbp]
    \centering
    \includegraphics[height = 9.3cm, width = 8.65cm]{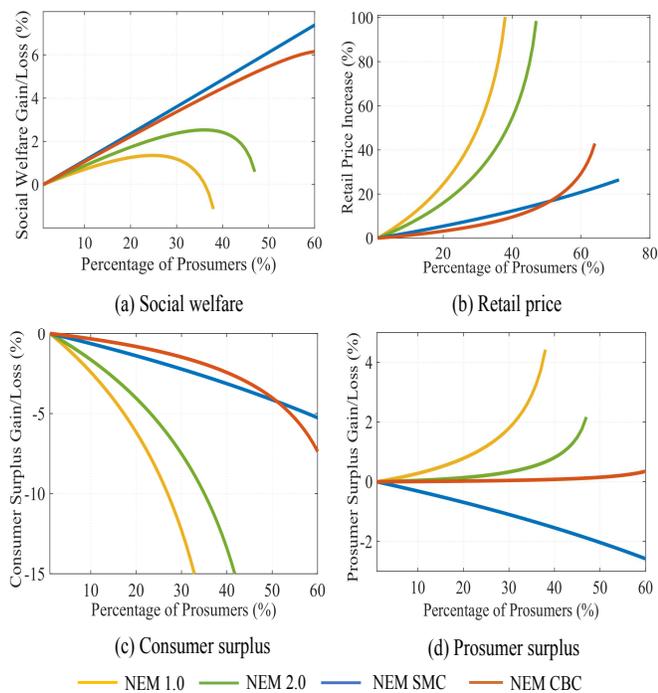}
    \caption{Normalized social welfare and customers surpluses gain/loss.}
    \label{fig:welfare}
\end{figure}

The social welfare was directly calculated using (\ref{eq:welfare}).  Fig.~\ref{fig:welfare} shows the percentage change (over the cases with 0\% adoption rate) in the total social welfare, the prosumers and consumers surpluses, and the retail prices as functions of the prosumer population size $\gamma$.

\par Fig.~\ref{fig:welfare} (a) shows the normalized social welfare of the four studied policies. In all cases, we observed increasing social welfare until the prosumer population exceeded a certain threshold, beyond which the social welfare started to decline.   NEM 1.0 (yellow) showed the earliest decline in social welfare, followed by NEM 2.0 (green).  The two successors NEM CBC  (red) and NEM SMC (blue) sustained the growth of social welfare at higher prosumer population levels. 
\par The rest of Fig. 4 explains the reasons that drove the social welfare downward when the adoption rate was high.  Fig.~\ref{fig:welfare} (b) shows the growing retail prices as the prosumer population increased. Recall that the retail price was the solution of the Ramsey pricing optimization where the revenue adequacy constraint was enforced.   NEM 1.0 and NEM 2.0 had the more generous bill savings for prosumers, resulting in lower revenue for the utility. Note that the price increases accelerated at the higher prosumer population, indicating potentials of pricing instability.

The retail rates of NEM CBC and NEM SMC increased modestly, showing that both policies effectively mitigated the revenue issues of NEM policies.   These two successor policies raised the revenue quite differently, however.  NEM CBC imposed the capacity-based connection charges on prosumers only, which raised additional revenue directly from prosumers, effectively making prosumers pay part of the avoided operating costs.  NEM SMC, on the other hand, reduced sell rate to its minimum, putting DER-exporting prosumers on an approximately equal footing as wind/solar farms participating in the wholesale electricity market. The sharper uptick of retail prices under NEM CBC at high adoption rate is due to the fact that price differential between retail and sell rates were kept the same for all adoption levels. In contrast, the gap between the two rates under NEM SMC increased with the DER adoption rate.

\par Fig.~\ref{fig:welfare} (c-d) show the consumer (non-adopters) and prosumer surpluses, respectively, all directly affected by the increasing retail (and sell) rates in Fig.~\ref{fig:welfare} (b).  Because of the concavity of consumption utilities and the linearity of the consumption cost, the retail price increases dominated the consumer surpluses, resulting in the downward trend of all consumer surpluses in Fig.~\ref{fig:welfare} (c). For prosumers, except under NEM SMC, the combined effects of increasing sell rate (as a result of retail rate increase) and reduced net consumption resulted in the overall increase of prosumer surplus.  Because NEM SMC had a fixed sell rate at the social marginal cost, the increasing retail rate became the dominant factor that drove the prosumer surplus down. Recall that the total social welfare included the social/health benefits of DER integration, which grew linearly with the prosumer population.  It was this linear growth that reversed the downward trends of consumer and prosumer surpluses under SMC.
\par The consistent but slowly growing retail rate under NEM SMC resulted in small  differences between the prosumer and consumer surpluses.  Together with the environmental and health benefits of BTM DER, NEM SMC yielded a higher social welfare compared to other policies. However, the effect of NEM SMC in reducing the gap between the consumer and prosumer surpluses has the undesirable consequence of disincentivzing DER adoption.  Sec.~\ref{subsub:num_PaybackTime}.

\subsection{Cost-shifts from prosumers}
\begin{figure}[htbp]
    \centering
    \includegraphics[scale=0.55]{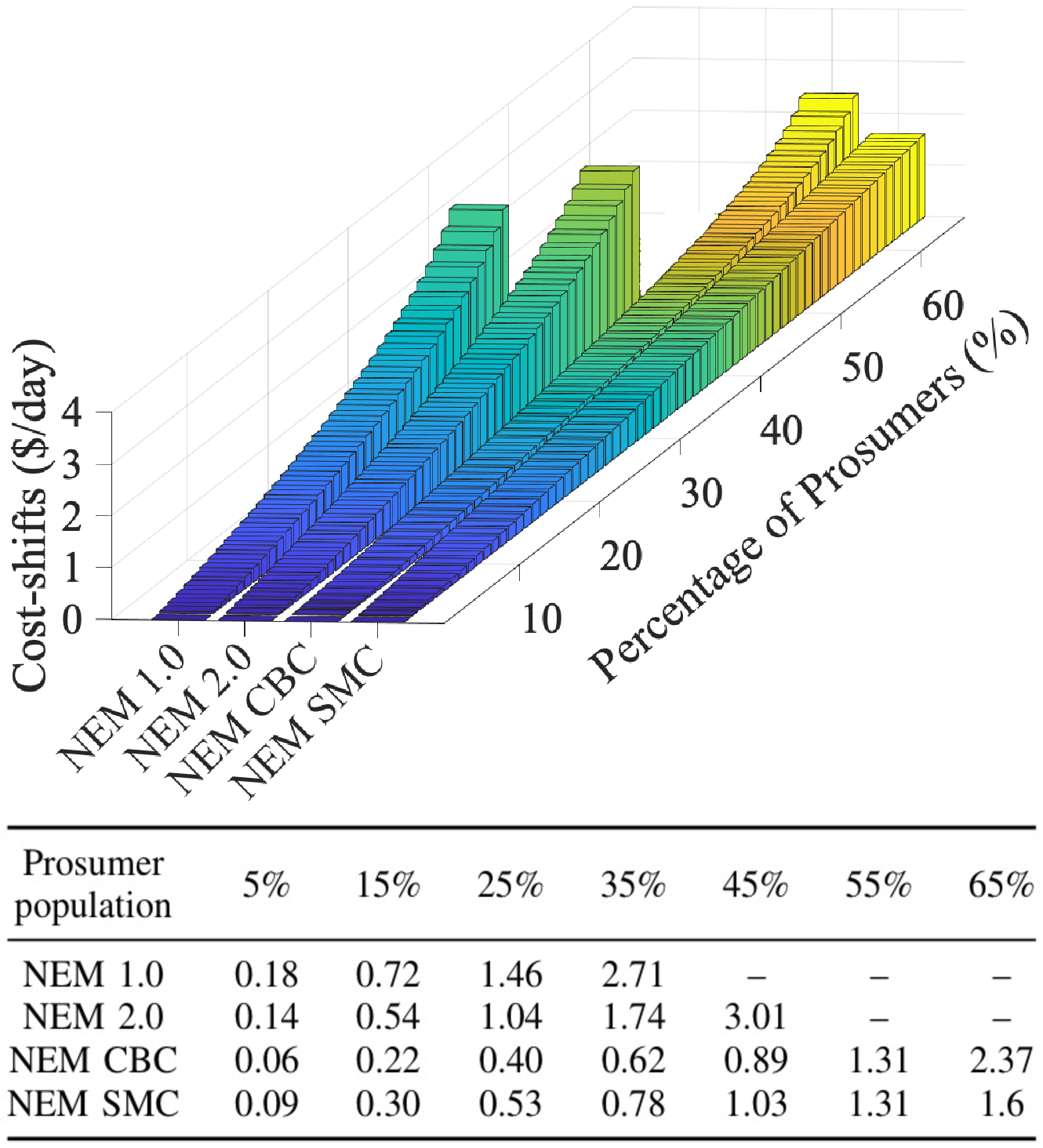}
    \caption{Cost-shifts from prosumers.}
    \label{fig:costshift}
\end{figure}

Current discussion of reforming NEM policies is  motivated in part by the cross-subsidies of the prosumer by consumers. Using the expected cost-shift (14) as a measure of cross-subsidy, Fig.~\ref{fig:costshift} shows the level of cost-shift (\$/day) against prosumer population ($\gamma$) under the four tariff structures.  As expected, NEM 1.0 had accelerated cost-shifts (cross-subsidies) because of its high compensation rate to the prosumer.  By reducing the sell rate, NEM 2.0 reduced cross-subsidies.  Our simulation also showed that the current Californian NEM 2.0 (one-part TOU) tariff was more effective in reducing subsidies than the two-part tariff NEM 1.0 with up to \$11/month connection charge. Note that the increase of the retail rate with adoption popularion under NEM 1.0 and NEM 2.0 (Fig.~\ref{fig:welfare} (b)) resulted in an infeasible break-even condition at 38\% and 47\% adoption rates under NEM 1.0 and NEM 2.0, respectively.

The cost-shifts were further lowered under the two NEM 2.0 successor policies.  NEM SMC, on average, yielded 92\% reduction of cost-shifts compared to NEM 1.0 and 88\% compared to NEM 2.0. NEM CBC was the most effective in suppressing subsidies, as it had a twofold effect on cost-shifts: a) the reduction of the retail price under the revenue adequacy condition, and b) the reduction of bill-saving through non-volumetric charge.  Although NEM CBC is more robust in achieving lower subsidies during low adoption levels, NEM SMC, as shown in Fig.~\ref{fig:costshift}, became more effective when the retail price increased given the unbundling of retail and sell rates under this policy. NEM SMC has already been applied in some states \cite{dsireArizona}, and has been proposed by many researchers as an equitable and cross-subsidy reducing compensation design \cite{Borenstein_canNetMetering}.

\begin{figure}[htbp]
    \centering
    \includegraphics[scale=0.415]{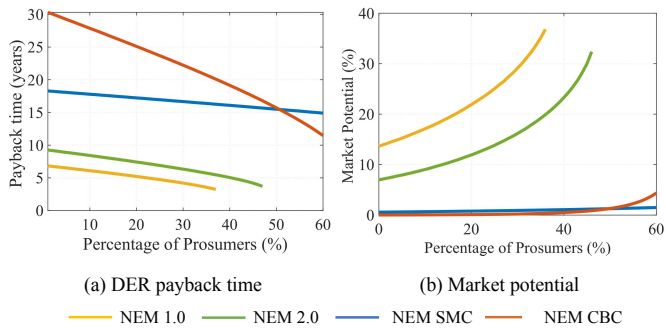}
    \caption{System payback time and market potential\protect\footnotemark. }
    \label{fig:payback}
\end{figure}

\subsection{Payback time and market potential}
\label{subsub:num_PaybackTime}

The DER investment payback time and market potential was calculated according  (\ref{eq:TNPpaybackTime}) and  \cite{denholm_drury_margolis_2009}, respectively.
Fig.~\ref{fig:payback} shows the expected payback time and the underlying market potential for the policies in table \ref{tab:Plotcasesarxiv}.  Fig.~\ref{fig:payback} shows that (a)  the transition in California from NEM 1.0 (yellow) to NEM 2.0 (green) prolonged the payback time by more than 45\%, on average. As a result, the market potential in Fig.~\ref{fig:payback} (b) was reduced under NEM 2.0, by over than 5\%. Both NEM CBC (red) and NEM SMC (blue) greatly increased the payback time and suppressed market potentials of DER adoption, which is consistent with their more aggressive approaches in reducing surpluses from BTM DER contribution to the grid.  NEM CBC had a more dramatic increase of payback time because it penalized DER adopters with heavy connection charges since it does not affect the social welfare objective. As the prosumer population increases, the revenue from CBC grew linearly and cost of system operation in the short-run analysis stayed the same. The revenue break-even condition resulted in a significant decline in payback time with increasing prosumer population size as shown in Fig.~\ref{fig:payback} (a). The underlying dynamics that caused the significant decline of CBC payback time did not exist in NEM SMC.

Although the market potential is not a very accurate estimate of the actual rate of DER adoption, Fig.~\ref{fig:payback} (b) does suggest that the eventual adoption rate under  NEM 2.0 would be lower than that under NEM 1.0. The effects of the two successors on market potential were more substantial. If these policies are not changed over the years, the DER penetration will remain very low.
\par It is important to note that the short-run analysis of payback time and market potential are inadequate in characterizing the much more complex adoption dynamics where population size, tariff parameters, and technology parameters (such as PV costs) are endogenously determined. See \cite{8675474}.

\footnotetext{The system's degradation factor $\nu=0.5\%$, and the currency inflation rate  $\zeta=2.4\%$ data were taken from \cite{drury_denholm_margolis_2011}. For market potential, a 0.2 payback sensitivity and 90\% market size were used as in\cite{denholm_drury_margolis_2009}.}

\section{Discussion and Future Work} \label{sec:discussion}
\subsection{Beyond NEM X:  fit-in tariff}
The essential characteristics of NEM is that the total payment of a prosumer is determined by the {\em net energy consumption/production}. Under FiT \cite{zinaman_aznar}, however, separate meters are used to measure BTM DER production and prosumer consumption, and the total payment is computed using  data from both meters.  In this context, FiT does not belong to the NEM tariff family.  However, the methodology developed in this work applies to FiT. The prosumer decision problem becomes simpler under FiT because consumption and BTM DER generation are decoupled.  Direct comparison between FiT and NEM, however, is problematic because the two classes of tariffs realize incentives very differently.

\subsection{Uncertainty in BTM renewables and BTM storage}
 For the prosumer decision problem, we assume that the BTM renewables are  known within each billing period, which is reasonable when the billing period is short.  A more accurate model is to treat renewables as random processes and solve a stochastic decision problem. Characterizing the optimal prosumer decision policy is nontrivial and outside the scope of this work.  It is, however, not difficult to incorporate the  closed-form solution in Theorem~\ref{thm:structure} to develop a model predictive control strategy using forecasted renewables \cite{AlahmedTong_DERA:21}.

Note that uncertainties in BTM renewables are not entirely ignored in this work.  In particular, we model explicitly the randomness in prosumer's decision in the regulator's rate setting process in Sec \ref{sec:short} and develop a scenario-based stochastic Ramsey pricing model.

Deploying BTM storage together with PV can improve efficiency and reliability of renewable integration.  The model considered in this paper does not include storage explicitly, although our results apply to cases when the BTM storage is operated independently to modify available renewables. The joint optimization of storage operation and prosumer consumption is more challenging and is considered in  \cite{AlahmedTong_DERA:21}.

\section{Conclusion}\label{sec:conclusion}
NEM is one of the most significant pricing policies in the retail market for integrating BTM DER at scale.  This work represents a first attempt to establish an analytical and computational framework in analyzing and comparing NEM policy choices.  By obtaining a closed-form characterization of the optimal prosumer consumption policy, we provide insights into the rational consumption decisions that prioritize different types of consumptions based on the utilities of the consumption and the level of DER production. The short-run analysis of NEM policies sheds light on the tension between minimizing cost-shifts and market potentials.

\par While the proposed NEM X tariff model applies broadly to many rate-setting practices, the results presented here are limited in several aspects.  For instance, the prosumer decision model considered here does not explicitly capture stochasticities of renewables, and the joint optimization of consumption and storage operations is not considered.   Also missing is the long-run analysis of NEM X,  where a dynamic rate-setting process similar to that presented in \cite{8675474} needs to be considered.  The developed NEM X model and insights gained in the analysis are stepping stones toward a more complete understanding of a highly complex engineering economic problem.

{
\bibliographystyle{IEEEtran}
\bibliography{NEM_Tong}
}

\section*{Appendix A: Proofs}
\subsection*{Proof of Theorem~\ref{thm:structure}}
Recall the prosumer optimization under A1-A3:
\beq
 \begin{array}{lll}
\Pc: &  \underset{\dbf  \in \mathbb{R}^M}{\rm minimize}&P^{\pi}\big(\sum_{i=1}^M d_i - r\big) -  \sum_{i=1}^M U_i(d_i)\\
 & \mbox{subject to} & \bar{d}_i \ge d_i \ge 0,~~\forall i.\\
 \end{array}
 \label{eq:prosumerOPT1}
 \eeq
 Note that $\Pc$ is convex with a non differentiable objective.

We break the above optimization into three convex optimizations, $\Pc^+, \Pc^-$ and $\Pc^o$, corresponding to the three scheduling zones in Theorem~\ref{thm:structure}:
 \bea
& & \begin{array}{lll}
\Pc^+:& \underset{\dbf  \in \mathbb{R}^M}{\rm minimize}&  \pi^+ (\sum_{i=1}^M d_i -r ) - \sum_{i=1}^M U_i(d_i)  \\
 & \mbox{subject to} & \bar{d}_i \ge d_i \ge 0,~~\forall i,\\
 & &   \sum_i d_i - r  \ge 0. \\
 \end{array}
 \label{eq:P+}\\[0.5em]
& & \begin{array}{lll}
\Pc^-:&  \underset{\dbf  \in \mathbb{R}^M}{\rm minimize}& \pi^-(\sum_{i=1}^M d_i -r )  -  \sum_{i=1}^M U_i(d_i)    \\
 &  \mbox{subject to} & \bar{d}_i \ge d_i \ge 0,~~\forall i,\\
&  &   \sum_i d_i - r  \le 0. \\
 \end{array}
 \label{eq:P-}\\[0.5em]
& & \begin{array}{lll}
\Pc^o:&  \underset{\dbf  \in \mathbb{R}^M}{\rm minimize}& -  \sum_{i=1}^M U_i(d_i)   \\
 & \mbox{subject to} &  \bar{d}_i \ge d_i \ge 0,~~\forall i,\\
 & &    \sum_i d_i - r  = 0. \\
 \end{array}
 \label{eq:Po}
 \eea
 Given $r$, the optimal schedule is the one that achieves the minimum value among $\Pc^+, \Pc^-$ and $\Pc^o$.  Note that, for all three optimizations, optimal consumptions exist.  Because  the Slater's condition is satisfied for these optimizations,
KKT conditions for optimality is necessary and sufficient.

 We prove Theorem~\ref{thm:structure} with Lemma~\ref{lem1}-\ref{lem2}.

 \begin{lemma}[Schedule in the net production and consumption zones]  \label{lem1}
It is optimal to consume  $(d_i^+)$  when $r<d^+$ and $(d^-_i)$   when $r>d^-$.
 \end{lemma}

 {\em Proof:}    First, we show that, if the prosumer is to consume when $r<d^+$, it is optimal to consume with $(d^+_i)$.

 Under $\Pc^+$, the Lagrangian $\Lc^+$ is given by
\begin{align*}
\Lc^+  &= (\pi^+-\mu^+)  \bigg(\sum_{i=1}^M d_i -r \bigg)   - \sum_i \lambda_i^+ d_i\nn \\& +\sum_i \gamma_i^+ \left(d_i-\bar{d}_i \right)  -\sum_{i=1}^M U_i(d_i),\nn
\end{align*}
where $\mu^+, \lambda^+_i,\gamma^+_i \ge 0$ are Lagrange multipliers for the net-consumption constraint, the lower and upper consumption limit constraints, respectively.  We simply have to check that $(d_i^+)$ defined in (\ref{eq:d_i^+})
\[d_i^+ = \max\{0, \min\{V_i^{-1}(\pi^+),\bar{d}_i\}\},~~\forall i, \]
satisfies the KKT condition with properly chosen $(\mu, \lambda_i^+, \gamma_i)$.   First, if $r < d^+ = \sum_i d_i^+$, we must have $\mu=0$.  
Next, for $(d_i^+,\lambda_i^+, \gamma_i^+)$ to satisfy the KKT condition, we have
\begin{align*}
V_i(d_i^+) &= \pi^+ - \lambda_i^++\gamma^+_i~ \Rightarrow~ d^+_i = V_i^{-1}(\pi^+ - \lambda^+_i+\gamma^+_i). \label{eq:KKTproof}
\end{align*}
If $0 \le V^{-1}_i(\pi^+) \le \bar{d}_i$, then $$(d_i^+=V_i^{-1}(\pi^+), \gamma_i^+=0, \lambda_i^+=0)$$ satisfies the part of  KKT condition involving device $i$.  Therefore, $d_i^+=V_i^{-1}(\pi^+)$ is optimal for device $i$'s consumption.

If $V^{-1}_i(\pi^+) > \bar{d}_i$, the monotonicity of $V^{-1}$ implies that we can find $d^+=\bar{d}_i$,  $\lambda_i^+=0$ and some $\gamma_i^+ > 0$ satisfying the KKT condition.  Therefore,   $d_i^+=\bar{d}_i$ is optimal. Likewise, if $V^{-1}_i(\pi^+) < 0$, we must have $d_i^+=0$.

In summary,  the optimal consumption for device $i$ is
\[d_i^+ = \max\{0, \min\{V_i^{-1}(\pi^+),\bar{d}_i\}\},~~\forall i. \]

 %Therefore,
%\begin{align*}
%d_i^+ &= \left\{\begin{array}{lll}
%V^{-1}_i(\pi^+) & \gamma_i^+=0, \lambda_i^+=0\\
%V^{-1}_i(\pi^++\gamma_i^+)&  \gamma_i^+>0, \lambda_i^+=0\\
%V^{-1}_i(\pi^+-\lambda_i^+) &  \gamma_i^+=0, \lambda_i^+>0\\
%\end{array}
%\right.\nn\\&=
%\left\{\begin{array}{lll}
%V^{-1}_i(\pi^+) & \gamma_i^+=0, \lambda_i^+=0\\
%\bar{d}_i&  \gamma_i^+>0, \lambda_i^+=0\\
%0 &  \gamma_i^+=0, \lambda_i^+>0\\
%\end{array}
%\right..
%\end{align*}
%Note that since $\bar{d}_i \neq 0, \forall i$, the case $\gamma_i^+,\lambda_i^+>0$  can be eliminated.

%The optimal consumption of device $i$ is therefore $d_i^+$.

Similarly, when $r>d^-$, the optimal consumption of device $i$ is shown to be:
\[d_i^- = \max\{0, \min\{V_i^{-1}(\pi^-),\bar{d}_i\}\},~~\forall i, \]
where, from the monotonicity of $V^{-1}$, $d^-_i \geq d^+_i, \forall i$. $\blacksquare$

 \begin{lemma}[Schedule in the net-zero zone] \label{lem2}
 When $d^+\le r \le d^-$,  it is optimal to match the consumption to $r$ with schedule $(d^o_i(r))$ where $d_i^o(r)$ is continuous and monotonically increasing function of $r$ in $[d^+,d^-]$.
 \end{lemma}

 {\em Proof:} First, we show that, if the prosumer is to be a zero net energy consumer, it is optimal to schedule with $(d_i^o)$.

 Under $\Pc^o$, the Lagrangian is given by
 \begin{align*}
\Lc^o &= \mu^o (\sum_{i=1}^M d_i -r )   - \sum_i \lambda_i^o d_i\nn\\& +\sum_i \gamma_i^o \left(d_i-\bar{d}_i \right)  -\sum_{i=1}^M U_i(d_i),~~\lambda_i^o, \gamma_i^o \ge 0.\nn
\end{align*}
By the KKT condition, the optimal schedule $d_i^o$ and the associated Lagrange multipliers $\mu^o,\lambda_i^o, \gamma_i^o\geq 0$ must satisfy
\[
V_i(d_i^o) = \mu^o - \lambda_i^o+\gamma_i^o.
\]
%Solving the above equation, we have
%\begin{align*}
%d_i^o &= \left\{\begin{array}{lll}
%V^{-1}_i(\mu^o) & \gamma_i^o=0, \lambda_i^o=0\\
%V^{-1}_i(\mu^o+\gamma_i^o)&  \gamma_i^o>0, \lambda_i^o=0\\
%V^{-1}_i(\mu^o-\lambda_i^o) &  \gamma_i^o=0, \lambda_i^o>0\\
%\end{array}
%\right.\nn\\&=
%\left\{\begin{array}{lll}
%V^{-1}_i(\mu^o) & \gamma_i^o=0, \lambda_i^o=0\\
%\bar{d}_i&  \gamma_i^o>0, \lambda_i^o=0\\
%0 &  \gamma_i^o=0, \lambda_i^o>0\\
%\end{array}
%\right.\nn\\
%&= \max\{0, \min\{V_i^{-1}(\mu^o),\bar{d}_i\}\},
%\end{align*}
Using the same argument used in the proof of Lemma 1,
\[
d_i^o= \max\{0, \min\{V_i^{-1}(\mu^o),\bar{d}_i\}\},
\]
where $\mu^o$ must be such that the equality constraint holds:
\beq
\sum_{i=1}^M \max\{0, \min\{V_i^{-1}(\mu^o),\bar{d}_i\}\} = r.
\label{eq:equality}
\eeq

When $r>d^-$, we show, similarly,  the optimal consumption of device $i$ is  $d_i^-$.

Next, we show that (\ref{eq:equality}) must have a non negative  solution when $d^+\le r\le d^-$,  let
\[
F(x):= \sum_{i=1}^M \max\{0, \min\{V_i^{-1}(x),\bar{d}_i\}\} -r.
\]
Note that $F(\cdot)$ is continuous and monotonically decreasing.  Because
 \[
 F(\pi^+) \le 0,~~F(\pi^-) \ge 0,
 \]
there must exists $\mu^o \in [\pi^-,\pi^+]$ such that $F(\mu^o)=0$. Therefore, (\ref{eq:equality}) must have positive solution, which also implies that
\begin{equation}
\label{eq:consumptionInequalityProof}
    d_i^+ \le d_i^o(r) \le d^-_i.
\end{equation}

Furthermore, the continuity and monotonicity of $F$ in $r$  implies that $d_i^o(r)$ is continuous and monotonically increasing function of $r$.  $\blacksquare$

\subsection*{Proof of Theorem~\ref{thm:comparative}}  We prove Theorem~\ref{thm:comparative} in two steps.
\subsubsection{Monotonicity in $r$}
To show the monotonicity of $(d_i^\pi(r))$,  we note that the monotonicity of $V_i$ and $\pi^+ \ge \pi^-$ imply that $d^-_i  \ge  d^+_i$. Within the net-zero zone, from Theorem~\ref{thm:structure},
$d^+_i  \le d_i^o(r) \le d^-_i$. By Lemma~\ref{lem2}, $d_i^o(r)$ is continuous and monotonically increasing in $[d_i^+, d_i^-]$.  Thus $d_i^\pi(r)$ is continuous and monotonically increasing for all $r>0$.

Next, to show that the total payment $P^\pi(\sum_i^M d^\pi_i(r) -r)$ decreases monotonically  with $r$, we note that  (i) the optimal scheduling of the prosumer is such that, as $r$ increases,  it changes from the consumption zone to the net-zero zone to the production zone; (ii) the payment from the prosumer to the utility is only positive in the production zone.

Finally, to show that the prosumer surplus $S^\pi(r)$  is monotonically increasing, we note that
\[
S^\pi(r) = \sum_i  U_i(d_i^\pi(r)) - P^\pi\big(\sum_i d^\pi_i(r) -r\big).
\]
Because $U_i$ is monotonically increasing with $d_i^\pi(r)$ therefore with $r$,  and the payment is monotonically decreasing with $r$, the prosumer's surplus is monotonically increasing with $r$.

\subsubsection*{Monotonicity in $\pi$}   Consider first the monotonicity with respect to $\pi^+$ at a fixed $r$.
From the definition of $d^+$ and the monotonicity of $V^{-1}_i$,  $d^+ \downarrow$ as $\pi^+ \uparrow$, and $d^-$ stays unchanged.

Note that $\pi^+$ only influences $d_i^\pi(r)$  in the consumption zone defined by $r  <  d^+$, within which
$d_r^\pi(r) = \max\{0, \min\{V_i^{-1}(\pi^+),\bar{d}_i\}\}$. Assumption A2  implies that $d_r^\pi(r) \downarrow$ as $\pi^+ \uparrow$.
For $r=d^+$ and  every $\tilde{\pi}\ge \pi^+$, $\tilde{r}^+:=\sum_i \max\{0,\min\{V_i^{-1}(\tilde{\pi}),\bar{d}_i\}\} < d^+=r$. Therefore, the prosumer is in the net-zero zone in which the consumption level is independent of $\pi^+$.  Therefore, $d_i^{\tilde{\pi}}(r)=d_i^\pi(r)$.  Thus we have  $d_r^\pi(r)$ is monotonically decreasing in $r$.

Now consider the prosumer surplus $S^{\pi}(r)$ for fixed $r$.   Suppose that $r<  d^+$, by the envelope theorem,
\bea
\frac{\partial}{\partial \pi^+} (-S^{\pi}(r)) &=& \frac{\partial}{\partial \pi^+} \Lc^+ \nn\\
&=& \big(\sum_i d_i^\pi(r) -r\big) >0,\nn
\eea
where $\Lc^+$ is the Lagrangian of optimization $\Pc^+$.   Therefore, as $\pi^+ \uparrow$, the objective of $\Pc^+$ increases, and total surplus decreases.  In the net-zero and production zones,  $S^\pi(r)$ is not a function of $\pi^+$; thus $S^\pi(r)$ is unchanged with respect.in the consumption zone and the size of the consumption zone.

Next, we consider the monotonicity with respect to $\pi^-$, which has influence on the endogenous variables only in the production zone.  When $r > d^-$, $d_i^\pi(r)=d^-_i$.   From (\ref{eq:d_i^-}), we have $d_i^\pi(r) \downarrow$ as $\pi^- \uparrow$.

For the prosumer surplus when $r > d^-$, we have
\bea
\frac{\partial}{\partial \pi^-} (-S^{\pi}(r)) &=& \frac{\partial}{\partial \pi^-} \Lc^- \nn\\
&=& \big(\sum_i d_i^\pi(r) -r\big) < 0,\nn
\eea
where $\Lc^-$ is the Lagrangian of optimization $\Pc^-$. Therefore, as $\pi^- \uparrow$, the objective of $\Pc^-$ decreases, and total surplus increases.

For the prosumer payment $P^\pi(r)$ when $r > d^-$, we have
\[
P^\pi(r) = - \pi^- ( r- \sum_i d_i^{\pi} (r)) \le 0.
\]
Because the consumption decreases as $\pi^+$ increases, we have $P^\pi(r) \downarrow$  (becoming more negative).

Finally, the monotonicity of $\pi^0$ is immediate from the definitions.  $\blacksquare$

\subsection*{Proof of proposition~\ref{prop:loadranking}}
We utilize the monotonicity of $V^{-1}$, and the price inequality $\pi^+ \geq \pi^-$, which together imply $d^-_i \geq d^+_i$. The proof of each case is in order:
\begin{enumerate}
    \item If $V_i(0) > \pi^+$,  because $V_i(0)$ is a monotonically decreasing function, $d_i^+ = \min\{V_i^{-1}(\pi^+),\bar{d}_i\}>0$.  From (\ref{eq:consumptionInequalityProof}), we have $d^-_i \geq d^o_i(r)\geq  d^+_i>0$.  Therefore, device $i$ will always scheduled to consume in one of the three zones.
    \item If $V_i(0) \le \pi^+$, then $V^{-1}(\pi^+)<0$. From the proof of lemma \ref{lem1}, we must have $\gamma_i^+=0$, and $\lambda_i^+>0$. Therefore, $d_i^+=0$;  it is optimal not to schedule the device in the net consumption zone.
    \item If $ V_i(0) < \pi^-\le \pi^+$, then $V^{-1}(\pi^-)<0$.  From the KKT condition,  we must have $d_i^-=d_i^+=0$. It is optimal not to use device $i$ in all scenarios.
\end{enumerate}
\hfill $\blacksquare$

\section*{Appendix B: Utility Parameters Setting}
\label{appendix:UtilityParameters}
We adopt a widely-used quadratic concave utility function of the form:
\begin{equation}
    \label{eq:example-utility-function}
    U_i(d_i) = \alpha_i d_i - \frac{1}{2}\beta_i d_i^2, 
\end{equation}
where $\alpha_i, \beta_i$ are some utility parameters that are dynamically calibrated.
\par Three load types with three different utility functions of the form in (\ref{eq:example-utility-function}) were considered: 1) HVAC load\footnote{\label{HVAC} The residential load profile data is taken from NREL open dataset for a nominal household in Los Angeles. We used the summer months data, that is June-August, 2019. The data can be found at: \url{https://openei.org/datasets/files/961/pub/RESIDENTIAL_LOAD_DATA_E_PLUS_OUTPUT/HIGH/USA_CA_Los.Angeles.Intl.AP.722950_TMY3_HIGH.csv}.}, 2) EV load\footnote{The EV load data is taken from NREL EV Infrastructure Projection (EVI-Pro) simulation tool for the city of Los Angeles, CA: \url{https://afdc.energy.gov/evi-pro-lite/load-profile}}, 3) other household loads such as lighting and appliances\textsuperscript{\ref{HVAC}}.
As introduced in \cite{campaigne_balandat_ratliff_2016}, the historical retail prices\footnote{We use historical PG\&E prices, which can be found at: \url{https://www.pge.com/tariffs/electric.shtml}.} and historical consumption data are used to calibrate the quadratic utility function parameters by predicating an elasticity of demand\footnote{The HVAC and household appliances elasticity values are taken from \cite{ASADINEJAD201826}, and the EV charging elasticity value is taken from \cite{doi:10.1086/689702}}. Considering that for historical data, the price differential is zero, then an interior solution of $(\ref{eq:prosumerOPT})$ yields:
\[ d_i(\pi^h) = \frac{\alpha_i-\pi^h}{\beta_i},
\]
where $\pi^h$ is the historical retail price. For each load type $i$ having an elasticity of demand $\varepsilon_i$, the elasticity can be expressed as:
\[
\varepsilon_i (\pi^h) = -\frac{\partial d_i^h(\pi^h)}{\partial \pi^h} \frac{\pi^h}{d_i^h}  = -\frac{1}{\beta_i}\frac{\pi^h}{d_i^h} =\frac{\pi^h}{\alpha_i-\pi^h} .
\]
Solving for $\alpha_i$ and $\beta_i$, we get:
\[\alpha_i = -  \left( \frac{1-\varepsilon_i}{\varepsilon_i}\right)\pi^h \]
\[\beta_i = -\frac{\pi^h}{\varepsilon_i d_i^h}.\]
$\alpha_i$ and $\beta_i$ are calibrated for each time period based on the realized prices and consumption data.

\begin{comment}

\section*{Appendix C: Case Studies}
\label{appendix:CaseStudies}
\input appendix_case_studies_R1
\end{comment}

\end{document}